# Ultrafast photocurrents and THz generation in single InAs-nanowires

Nadine Erhard[1], Paul Seifert[1], Leonhard Prechtel[1], Simon Hertenberger[1], Helmut Karl[2], Gerhard Abstreiter[1,3], Gregor Koblmüller[1], and Alexander W. Holleitner[1]

1    Walter Schottky Institut and Physik-Department, Technische Universität München, Am Coulombwall 4a, 85748 Garching, Germany
2    Institute of Physics, University of Augsburg, 86135 Augsburg, Germany
3    Institute for Advanced Study, Technische Universität München, Lichtenbergstraße 2, 85748 Garching, Germany

**To clarify the ultrafast temporal interplay of the different photocurrent mechanisms occurring in single InAs-nanowire-based circuits, an on-chip photocurrent pump-probe spectroscopy based on coplanar striplines was utilized. The data are interpreted in terms of a photo-thermoelectric current and the transport of photogenerated holes to the electrodes as the dominating ultrafast photocurrent contributions. Moreover, it is shown that THz radiation is generated in the optically excited InAs-nanowires, which is interpreted in terms of a dominating photo-Dember effect. The results are relevant for nanowire-based optoelectronic and photovoltaic applications as well as for the design of nanowire-based THz sources.**

Scanning photocurrent microscopy is a widely used method to investigate diffusive and ballistic carrier transport, and to map the electronic band bending in contacted nanostructures, such as semiconducting nanowires [1]-[10]. The photocurrent generation in single semiconducting nanowires has been explained either by carrier drift due to internal and external electric potentials [2], by a photo-thermoelectric effect [11],[12], or by carrier diffusion processes [10]. We recently presented an on-chip pump-probe photocurrent spectroscopy based on coplanar stripline circuits to investigate such photocurrent dynamics in single p-doped GaAs-nanowires with a picosecond time-resolution [13],[14]. Here, we investigate the ultrafast optoelectronic response of single nanowires made of InAs; a material which has been successfully utilized for bulk and planar THz detectors and emitters [15]-[20][18]. Recent work demonstrated the use of single InAs-nanowires as THz detectors [21]; and recent optical pump-probe experiments with a far-field detection scheme on ensembles of InAs-nanowires suggested that THz radiation is emitted from optically excited InAs-nanowires [22]. With our on-chip near-field detection circuitry, we verify and characterize the THz emission on the level of single InAs-nanowires. In particular, our data suggest that a dominating photo-Dember effect with the possibility of an additional optical rectification is the origin of the THz generation in such nanowires [15],[16],[19],[20],[23]. This interpretation is consistent with the fact that we observe the THz generation at all positions along the InAs-nanowires. At the metal contacts, an ultrafast photo-thermoelectric current as well as a displacement current are superposed to the THz generation. In addition, we verify the drift of photogenerated holes from all excitation positions along the InAs-nanowires to the metal contacts. We also compare our on-chip photocurrent pump-probe spectroscopy with a photocurrent autocorrelation technique. So far, the latter method has been applied to graphene-based [24],[25] and carbon nanotube-based circuits [26]. In the present case of the single InAs-nanowires, we observe that the ultrafast autocorrelation techniques yields time constants consistent with the time scales which we detect for the ultrafast photo-thermoelectric current and the drift of photogenerated holes. However, we also witness that the autocorrelation spectroscopy alone does not identify these processes unambiguously. The presented ultrafast optoelectronic results may prove useful for the design of photodetectors, photoswitches, solar cells, high-speed transistors, and THz sources based on InAs-nanowires.

We investigate catalyst free, nominally undoped InAs-nanowires, which are grown by molecular beam epitaxy (MBE) on a $SiO_x$ covered (111)-oriented Si substrate (growth rate ~0.24 Å/s, $As_4$ partial pressure ~1 x $10^{-5}$ mbar, growth temperature ~480 °C) [27],[28],[29]. The vertically aligned nanowires with a hexagonal cross section and {110} side-facets have a length of 5 μm and a diameter of ~130 nm. Fermi-level pinning at the (110)-nanowire surface leads to a n-type surface channel with a carrier concentration of $6 \times 10^{-11}$ $cm^{-2}$ [30].

The nanowires are mechanically transferred onto a preprocessed sapphire substrate in a random fashion and then contacted in a coplanar stripline circuit by optical lithography (strip width 5 μm, separation 4 μm, metal thickness Ti/Au 10 nm/150 nm) [14]. Figure 1a shows an optical microscopy image of such an electrically contacted InAs-nanowire. As depicted in Figure 1b, a field probe is positioned ~400 μm from the nanowire on a ~200 nm thick layer of MBE-grown silicon. Prior to the nanowire transfer, the silicon was $O^+$ ion-implanted with an energy of 90 keV at a fluence of $2 \times 10^{15}$ $cm^{-2}$ in order to reduce the charge carrier recombination lifetime in the silicon to below 1 ps [31]. Subsequently, the silicon was etched in an HF / $HNO_3$ dip to remain only at the predefined position of the field probe.

The circuitry sketched in Figure 1b allows an on-chip pump-probe detection of the time-resolved photocurrent $I_{Sampling}$. To this end, the InAs-nanowire in the stripline circuit is optically excited by a pump-pulse from a titanium:sapphire laser, polarized linearly along the nanowire axis, at a wavelength of $\lambda = 780$ nm, with a spot size of 2 μm and a bandwidth limited pulse duration of

~160 fs. After the excitation, an electro-magnetic pulse starts to travel along the stripline [32]. A sampling circuit senses the transient electric field of the travelling pulse at the position of the field probe (Figure 1b). Here, we utilize an Auston switch, based on the ion-implanted silicon on the sapphire substrate. The time delay between the pump- and probe-pulse is controlled by a delay stage. Measuring $I_{Sampling}$ in the sampling circuit as a function of the time delay yields information on the optoelectronic processes in the nanowire with a picosecond time-resolution [13],[14],[32]-[34].

Initially, we perform spatially resolved and time-integrated photocurrent measurements by scanning the pump laser with a scanning mirror across the nanowire and recording the current $I_{Photo}$ (see Figure 1b). Simultaneously to the photocurrent the reflectance is measured to deduce the position of the metal striplines, which are depicted as dashed lines in Figures 2a-c. Figures 2a, 2b, and 2c show photocurrent maps of $I_{Photo}$ for a bias voltage of $V_{SD}$ = -5 mV, 0 V, and +5 mV, respectively. At zero bias, a positive (negative) current $I_{Photo}$ is generated for excitation at the top (bottom) contact of the nanowire. At $V_{SD}$ = 5 mV (−5 mV), the positive (negative) current $I_{Photo}$ at the top (bottom) contact is enhanced whereas the negative (positive) current $I_{Photo}$ at the bottom (top) contact is suppressed [5],[7].

In order to identify the different mechanisms contributing to the time-integrated photocurrent $I_{Photo}$, we perform time-resolved photocurrent measurements. Figure 2d shows the time-resolved photocurrent traces $I_{Sampling}$ for different excitation positions along the nanowire starting from the top contact (triangle in Figures 2a-c, top trace in Figure 2d) to the bottom contact (square in Figures 2a-c, bottom trace in Figure 2d). Three features dominate the time-resolved photocurrent on different time scales. First, there is an initial peak up to 10 ps, which changes its polarity from the top to the bottom contact. Second, the first peak is superimposed by a fast oscillation of $I_{Sampling}$ during the first ~2 ps. Both features do not change in the examined range of

$|V_{SD}| \leq 8$ mV (data not shown). Third, for time scales longer than ~20 ps, a peak with a FWHM of several hundred picoseconds occurs in $I_{Sampling}$ (arrow). This peak shifts in time when the laser excitation is scanned from the top to the bottom contact. Furthermore, it changes its polarity with the applied bias voltage $V_{SD}$ (data not shown).

Assuming a clear time-scale separation of the different processes, the red lines in Figure 2d are fits to the data which consist of three components fitting the three features described:

$$I_{Fit}(t) = I_{Thermo}(t) + I_{THz}(t) + I_{Hole}(t) \qquad (1).$$

Figures 3a and b show the three fit components in detail taking the trace of position 3 in Figure 2d as an example. $I_{Thermo}(t)$ (dashed line) fits the initial peak with an exponentially modified Gaussian. $I_{THz}(t)$ (dotted line) consists of two Lorentzian peaks with different sign and amplitude to fit the oscillatory part of $I_{Sampling}$. The superposition of two Lorentzians allows us to get a good estimate of the area under the oscillatory contribution with positive and negative amplitudes (see section 3.2). The peak at 20-100 ps is fitted with

$$I_{Hole}(t) = I_0 |\exp(-(t-t_0)/\tau_{rise}) - \exp(-(t-t_0)/\tau_{decay})| \qquad (2),$$

(dashed dotted line) which can be used to fit the drift of photogenerated charge carriers [14]. $t_0$ defines the position of the peak and $\tau_{rise}$ ($\tau_{decay}$) is its rise-time (decay-time). In the following sections, the origin of the three different photocurrent components and the results of the fits are discussed in detail.

On the one hand, a localized non-resonant laser excitation of a semiconductor sample results in a local temperature increase of the corresponding electron and phonon baths, e.g. by relaxation of the photogenerated charge carriers via electron-electron scattering and phonon generation [35]. On the other hand, a temperature difference $\Delta T$ at the interface of two materials with different Seebeck coefficients $S$ can result in a thermoelectric current [12]:

$$I_{Thermo} = (S_{Nanowire} - S_{Stripline})\Delta T / R \qquad (3),$$

with $R$ the total resistance of the electrical circuit. The direction of $I_{Thermo}$ in (3) depends only on the sign of $\Delta T$, since in first approximation, $S_{Nanowire}$ and $S_{Stripline}$ are independent of the bias voltage. We observe that the first peak of $I_{Sampling}$ occurs predominantly at the nanowire-metal interface (e.g. positions 1-3 and 9-11 in Figure 2d) and it does not depend on $V_{SD}$ (data not shown). Both findings are strong indications for an ultrafast photo-thermoelectric current [14],[34]. In addition, both the form of the first peak – phenomenologically an exponentially modified Gaussian – and the observed time scale of $\tau_{Thermo}$ = (2-3) ps are consistent with recent results found on ohmic metal contacts of GaAs-nanowire-based circuits [14]. InAs-nanowires typically exhibit a surface electron density such that the nanowire-metal-interfaces turn out to be almost ohmic. We note that the time-integrated $I_{Photo}$ maps at zero $V_{SD}$ exhibit a maximum with opposite sign at each contact (e.g. Figure 2b). Such maxima are consistent with the interpretation of a photo-thermoelectric current. However, they are also consistent with the occurrence of Schottky contacts [5],[7] and in turn, an ultrafast dielectric displacement current [14]. Indeed, we see indications of such an ultrafast photocurrent at the nanowire-metal contacts in the data analysis presented in the next section 3.2.

The time-resolved $I_{Sampling}$ measured on InAs-nanowires exhibits a fast oscillation during the first couple of picoseconds for almost all positions along the nanowire (Figure 2d). We fit the oscillation by $I_{THz}(t)$ of equation (1) which comprises two Lorentzian functions with opposite sign (dotted line in Figure 3d). Figure 4a represents the same data as in Figure 2d but with the fitted photo-thermoelectric contribution subtracted, i.e. $I_{Sampling}(t) - I_{Thermo}(t)$. This analysis uncovers that the oscillatory behavior occurs all along the nanowire, and that the function $I_{THz}(t)$ nicely fits the oscillatory part. In a next step, we perform a fast Fourier Transformation (FFT) of the oscillatory data (open circles in Figure 4b) and fit the FFT amplitudes by the following expression (lines in Figure 4b):

$$A \cdot f \cdot \exp(-f/B) + C \quad (5),$$

with $f$ the frequency of the FFT. We observe FFT amplitudes up to a frequency of ~1 THz, which is the bandwidth of the coplanar stripline circuit on sapphire [14][34]. For the positions 1 to 8 along the nanowire, the FFT amplitudes have a maximum at $f_{max} = (0.30 \pm 0.04)$ THz. Only at position 9 to 11 along the nanowire, the maximum deviates from this value to be $f_{max} = (0.16 \pm 0.02)$ THz. This deviation is consistent with the observation in Figure 4a, that for position 11 the subtracted value $I_{Sampling}(t) - I_{Thermo}(t)$ yields just a peak with a FWHM of $(1.41 \pm 0.08)$ ps. The form of the latter photocurrent contribution and the time scale indicate the occurrence of an ultrafast dielectric displacement current at position 11, as recently observed at the position of Schottky contacts of GaAs-nanowires [14]. Hereby, $I_{Sampling}$ corroborates the existence of a Schottky contact at position 11. However, at position 1 of Figure 4a, there is no indication of a corresponding dielectric displacement current on top of the dominating photo-thermoelectric current.

Generally, the metal stripline circuit acts as a near-field antenna for THz radiation, greatly enhancing the out-coupling of electromagnetic fields in the direction of the stripline [34]. Hereby, we interpret the oscillatory part of the time-resolved $I_{Sampling}$ to stem from THz radiation generated in the optically excited InAs-nanowire. Several generation mechanisms can explain the generation of THz radiation in InAs-nanowires [22]: (i) an ultrafast acceleration of the photogenerated charge carriers caused by build-in electric fields [36], e.g. at the metal-nanowire interface, (ii) non-linear effects such as an optical rectification [17],[37], (iii) an ultrafast acceleration due to surface depletion fields [15],[38],[39], and (iv) a photo-Dember effect due to ambipolar diffusion of the photogenerated charge carriers [15],[16],[19],[20],[23]. The data in Figure 4a already indicate that the possibility (i) can be excluded, because the oscillatory photocurrent contribution of $I_{Sampling}$ is rather supressed than amplified at the metal-nanowire interface (positions 9 to 11). In order to identify the dominating mechanism out of possibilities (ii)-(iv), we perform a pump laser intensity dependence of the oscillatory part of $I_{Sampling}$ on a second InAs-nanowire at a position in between the corresponding two metal contacts (Figure 5a). In this nanowire sample, the photo-thermoelectric current is less dominant even at the contacts, such that the oscillatory part of $I_{Sampling}$ can be directly analyzed without the subtraction of the photo-thermoelectric contribution. We observe that the THz oscillation amplitude generally increases with excitation intensity (open circles in Figures 5a and 5b), and that the corresponding FFT amplitudes are again cut-off at a frequency of about 1 THz. We interpret this finding again such that the response function of the coplanar stripline circuit has a bandwidth of ~1 THz [14]. The data in Figures 5a and 5b are fitted by the same fit functions as in Figures 4a and 4b, respectively. We find that the area of the oscillatory part of $I_{Sampling}$ increases linearly with laser intensity (Figure 5c). Furthermore, we observe that $f_{max}$, the frequency at which the FFT amplitude is maximum (filled triangle in Figure 5b), is rather constant for a large part of the examined range of pump laser intensity (Figure 5d). A similar behavior has been reported for bulk n-doped GaAs-crystals [36], which was interpreted in terms of "cold electrons", confined within the surface depletion field of the GaAs-substrate which respond to the screening of the surface depletion field by the presence of photogenerated charge carriers. However, we tentatively exclude this possibility (iii) of an ultrafast acceleration of charge carriers due to surface depletion fields for the following two reasons. First, InAs is a narrow gap semiconductor and our n-type nanowires have a surface accumulation layer resulting in a small surface field [15],[22]. Second, similar time-resolved experiments on individual p-doped GaAs nanowires with a larger surface depletion field did not reveal THz emission [14]. Instead, we explain the THz-generation in the single InAs-nanowires by a photo-Dember effect based on the two following arguments. First, for small laser intensities, the Dember potential can be approximated to be proportional to the laser intensity which is consistent with Figure 5c [16],[40]. Second, although the area of the oscillatory part of $I_{Sampling}$ depends partially on the orientation of the linear polarization of the incident photons, we observe a polarization-independent offset (see

Figure 5e). This polarization-independent part of the THz oscillation is a clear indication for a photo-Dember effect [16]. The data in Figures 4a, 5a, 5b, 5c, and 5d are consistent with this conclusion. In our interpretation, the larger diffusivity of the photogenerated electrons compared to the one of the holes give rise to the photo-Dember effect. Generally, our circuitry is only sensitive to a coherent electromagnetic radiation [34], and the THz radiation is picked up by the coplanar striplines in a near-field regime. We explain the slight variation of the THz signal along the InAs-nanowire as in Figure 4a by the fact that the absorption is reduced at the contacts where the nanowire is covered by Ti/Au [41]. We note that the observed polarization dependence of the Figure 5e points towards an additional optical rectification at the {110} side-facets, as was shown for [110] surfaces of bulk n-InAs with zinc-blende structure [42]. However, our nanowires exhibit a significant part of wurtzite segments [27],[28],[29]. Furthermore, the dielectric confinement of the nanowires in combination with the metal contacts influences the polarization dependent absorption [41]. In turn, also the photo-Dember should exhibit a partial polarization-dependence as in the Figure 5e. We point out that D. V. Seletskiy et al. came to a similar interpretation of a dominating photo-Dember effect to explain their far-field data on an ensemble of InAs-nanowires [22]. There, the authors argue that a low energy acoustic surface plasmon mode of the nanowires allows coupling the THz radiation to the far-field radiation. In our case, we use a focused pump laser, which allows us to demonstrate that the THz radiation is emitted at all positions along the nanowires (Figure 4a).

At a time scale longer than ~20 ps, a broad photocurrent peak dominates $I_{Sampling}(t)$ (arrow in Figure 2d). We fit this peak by the third fit function component $I_{Hole}$, as described by equations (1) and (2). The peak shifts in time when the position of the pump pulse is scanned from the top to the bottom contact of the nanowire. In other words, the maximum of the peak defines the propagation time $t_{hole}$ which can be plotted versus the distance to the contacts (Figure 6) [13]. We can consistently fit this time-of-flight diagram with a line through the origin, which is the negatively biased top contact in Figures 2d and 6. In turn, we associate the third peak with the transport of photogenerated holes to the contacts. From the slope of the linear fit through the origin, we deduce the average velocity of the photogenerated holes to be $v_{\text{Hole}} = (2.9 \pm 0.3) \cdot$ ~~$10^7$ cm/s~~ $10^6$ cm/s. We numerically calculate the quantum velocity of the photogenerated holes for InAs to be on the order of $2.9 \times 10^6$ cms$^{-1}$ for both light and heavy holes at an excitation energy of 1.59 eV (supplementary material). Certainly, the overall dynamics of the photogenerated charge carriers is a combination of relaxation, thermalization, drift and diffusion dynamics [44], and the InAs-nanowires exhibit a significant part of wurtzite segments. Therefore, the calculation gives a rough estimate of the quantum velocity. For instance, the hole peak in Figure 2d broadens for excitation positions from the top to the bottom contact with a typical time scale of $\tau_{\text{rise}} = (10\text{-}60)$ ps, and $\tau_{\text{decay}} = (300\text{-}1000)$ ps. The broadening is consistent with an additional diffusion of photogenerated holes during their propagation to the contacts. The interpretation of the propagation of photogenerated holes as in Figure 6 are reproduced for $|V_{\text{SD}}| < 8$ mV (data not shown).

Finally, we note that the propagation of the photogenerated holes is the only bias dependent photocurrent component observed in the time-resolved $I_{\text{Sampling}}$ (Figure 2d). Consequently, the hole propagation apparently contributes considerably to the bias dependence of the time integrated photo-current $I_{\text{Photo}}$ (Figures 2a and 2c). In our time-resolved experiments, we do not observe any peak which can be explained in terms of a propagation of photogenerated electrons. Tentatively, we attribute this finding to the dominating photo-Dember effect all along the nanowires (Figure 4a) and a dominating electronic photo-thermoelectric effect at the contacts (Figure 2d).

Figure 7a shows the time-integrated $I_{\text{Photo}}$ versus the excitation intensity for the pump laser focused onto one of the nanowire-metal contacts (compare Figure 2b). $I_{\text{Photo}}$ saturates for intensities larger than ~20 kW/cm$^2$. We then apply a photocurrent autocorrelation technique by focusing both pump and probe pulse on the same position on the nanowire for an intensity in the saturation regime [24]-[26]. Figure 7b depicts the corresponding autocorrelation photocurrent versus the time-delay between pump and probe pulse. We can fit the data in Figure 7b by a tri-exponential decay function with the following time constants: $T_1$: $(1.8 \pm 0.1)$ ps, $T_2$: $(22 \pm 1)$ ps, and $T_3$: $(225 \pm 2)$ ps (given errors are fit errors). When we compare the values of $T_1$, $T_2$, and $T_3$ with time scales as discussed in sections 3.1, 3.2, and 3.3, we find the following consistencies. $T_1$ is consistent with the time scale of the photo-thermoelectric current or the displacement current. $T_2$ ($T_3$) is comparable to the $\tau_{\text{rise}}$ ($\tau_{\text{decay}}$) of the drift of the photogenerated holes. We interpret the obvious deviations between the time scales deduced from the two techniques from a certain uncertainty in the autocorrelation measurements. For instance, we observe that $T_1$, $T_2$, and $T_3$ depend on experimental details such that the pump and probe beams have the exact same excitation intensity, position, and spatial size [24]. Since the autocorrelation technique essentially works in the saturation regime, further measurements, such as temperature dependences, need to be undertaken to distinguish an ultrafast photo-thermoelectric current from a displacement current, as recently discussed for graphene-based circuits [25]. Our on-chip pump probe photocurrent spectroscopy, however, allows us to spatially resolve and thus characterize the various ultrafast photocurrent dynamics and to investigate the THz generation mechanism in nanoscale circuits [34].

In conclusion, we experimentally identify and separate the individual photocurrent mechanisms in single InAs-nanowires, which are overlaid and therefore individually not resolvable in time-integrated scanning photocurrent measurements. At the metal-nanowire interfaces, we identify ultrafast photocurrents which are interpreted in terms of a photo-thermoelectric current and a dielectric displacement current. At all positions along the nanowires, we observe the generation of THz radiation. The data suggest that a photo-Dember effect is the underlying process with the possibility of an additional optical rectification. For time scales longer than tens of picoseconds we observe a signal which is consistent with the drift dynamics of photogenerated holes. Our findings may prove essential for the design of ultrafast photoswitches, high-speed transistors, photodetectors, solar cells, and THz sources based on single InAs-nanowires.

We gratefully acknowledge financial support from the DFG, the German excellence initiative via the "Nanosystems Initiative Munich (NIM)", the Marie Curie FP7 Reintegration Grant and the ERC-2012-StG_20111012  project 306754.

**References**

[1]    J. Wang, M. S. Gudiksen, X. Duan, Y. Cui, and C. M. Lieber, Science **293**(5534), 1455-1457 (2001).

[2]    Y. Ahn, J. Dunning, and J. Park Nano Lett. **5**(7), 1367-1370 (2005).

[3]    H. Pettersson, J. Tragardh, A. Persson, L. Landin, D. Hessman, and L. Samuelson, Nano Lett. **6**(2), 229-232 (2006).

[4]    O. Hayden, R. Agarwal, and C. M. Lieber, Nature Mater. **5**, 352 (2006).

[5]    Z. Y. Zhang, C. H. Jin, X. L. Liang, Q. Chen, and L.-M. Peng, Appl. Phys. Lett. **88**(7), 073102 (2006).

[6]    J. E. Allen, E. R. Hemesath, and L. J. Lauhon, Nano Lett. **9**(5), 1903-1908 (2009).

[7]    S. Tunich, L. Prechtel, D. Spirkoska, G. Abstreiter, A. Fontcuberta i Morral, and A. W. Holleitner, Appl. Phys. Lett. **95**(8), 083111 (2009).

[8]    C. J. Kim, H.-S. Lee, Y.-J. Cho, K. Kang, and M. J. Jo, Nano Lett. **10**(6), 2043-2048 (2010).

[9]    M. E. Reimer, M. P. van Kouwen, M. Barkelid, M. Hocevar, M. H. M. van Weert, L. P. Kouwenhoven, V. Zwiller, R. E. Algra, E. P. M. Bakkers, M. T. Björk, H. Schmid, and H. Riel  J. Nanophot. **5**, 053502 (2011).

[10]   R. Graham, C. Miller, E. Oh, and D. Yu, Nano Lett. **11**(2), 717-722 (2011).

[11]   J. Cai, and G. D. Mahan, Phys. Rev. B **74**(7), 075201 (2006).

[12]   D. Fu, A. X. Levander, R. Zhang, J. W. Ager III, and J. Wu, Phys. Rev. B **84**(4), 045205 (2011).

[13]   L. Prechtel, S. Manus, D. Schuh, W. Wegscheider, and A. W. Holleitner Appl. Phys. Lett. **96**(26), 261110 (2010).

[14]   L. Prechtel, M. Padilla, N. Erhard, H. Karl, G. Abstreiter, A. Fontcuberta i Morral, and A. W. Holleitner, Nano Lett. **12**(5), 2337-2341 (2012).

[15] M. B. Johnston, D. M. Whittaker, A. Corchia, A. G. Davies, and E. H. Linfield, Phys. Rev. B **65**(16), 165301 (2002).

[16] P. Gu, M. Tani, S. Kono, and Kiyomi Sakai, J. Appl. Phys. **91**(9) 5533-5537 (2002).

[17] M. Reid, and R. Fedosejevs, Appl. Phys. Lett. **86**(1), 011906 (2005).

[18] G. D. Chern, E. D. Readinger, H. Shen, and M. Wrabeck Appl. Phys. Lett. **89**(14), 141115 (2006).

[19] K. Liu, Jingzhou Xu, Tao Yuan, and X.-C. Zhang, Phys. Rev. B **73**(15), 155330 (2006).

[20] M. Yi, K. Lee, J. Lim, Y. Hong, Y.-D. Jho, and J. Ahn, Opt. Express **18**(13), 13693-13699 (2012).

[21] M. Vitiello, L. Viti, L. Romeo, D. Ercolani, G. Scalari, J. Faist, F. Beltram, L. Sorba, A. Tredicucci, Appl. Phys. Lett. **100**(24), 241101 (2012).

[22] D. V. Seletskiy, M. P. Hasselbeck, J. G. Cederberg, A. Katzenmeyer, M. E. Toimil-Molares, F. Léonard, A. A. Talin, and M. Sheik-Bahae, Phys. Rev. B **84**(11), 115421 (2011).

[23] H. Dember, Phys. Z. **32**, 554 (1931).

[24] A. Urich, K. Unterrainer, and T. Mueller, Nano Lett. **11**(7), 2804-2808 (2011).

[25] D. Sun, G. Aivazian, A. M. Jones, J. S. Ross, W. Yao, D. Cobden, and X. Xu, Nat. Nanotechnol. **7**(2), 114–118 (2012).

[26] N. M. Gabor, Z. Zhong, K. Bosnick, and P. L. McEuen, Phys. Rev. Lett. **108**(8), 087404 (2012).

[27] G. Koblmüller, S. Hertenberger, K. Vizbaras, M. Bichler, J.-P. Zhang, and G. Abstreiter, Nanotechnology **21**(36), 365602 (2010).

[28] S. Hertenberger, D. Rudolph, S. Bolte, M. Döblinger, M. Bichler, D. Spirkoska, J. J. Finley, G. Abstreiter, and G. Koblmüller, Appl. Phys. Lett. **98**(12), 123114 (2011).

[29]  S. Hertenberger, D. Rudolph, J. Becker, M. Bichler, J. J. Finley, G. Abstreiter, and G. Koblmüller, Nanotechnology **23**(23), 235602 (2012).

[30]  T. D. Veal, and C. F. McConville, Appl. Phys. Lett. **77**(11), 1665-1667 (2000).

[31]  F. E. Doany, D. Grischkowsky, and C.-C. Chi, Appl. Phys. Lett. **50**(8), 460-462 (1987).

[32]  D. H. Auston, IEEE J. of QE, **19**(4), 639-648 (1983).

[33]  D. Krökel, D. Grischkowsky, and M.B. Ketchen, Appl. Phys. Lett. **54**(11), 1046 (1989).

[34]  L. Prechtel, L. Song, D. Schuh, P. Ajayan, W. Wegscheider, and A. W. Holleitner, Nat. Commun. **3**, 646 (2012).

[35]  J. Shah, Ultrafast Spectroscopy of Semiconductors and Semiconductor Nanostructures, Springer Series in Solid-State Sciences (Springer Berlin Heidelberg, 1999).

[36]  R. Kersting, K. Unterrainer, G. Strasser, H. F. Kauffmann, and E. Gornik, Phys. Rev. Lett. **79**(16),  3038-3041 (1997).

[37]  M. Bass, P. A. Franken, J. F. Ward, and G. Weinreich, Phys. Rev. Lett, **9**(11), 446-448 (1962).

[38]  X.-C. Zhang, and D. H. Auston, J. Appl. Phys. **71**(1), 326-338 (1992).

[39]  T. Dekorsy, T. Pfeifer, W. Kütt, and H. Kurz, Phys. Rev. B, **47**(7), 3842-3849 (1993).

[40]  W. Mönch, Semiconductor Surfaces and Interfaces, Springer Series in Surface Science, Vol. 26 (Springer Berlin, Heidelberg, New York, 1995).

[41]  S. Thunich, L. Prechtel, D. Spirkoska, G. Abstreiter, A. Fontcuberta i Morral, A.W. Holleitner, Appl. Phys. Lett. 95, 083111 (2009).

[42]  M. Reid. R. Fedosejevs, Proc. SPIE 5577, 659 (2004).

[43]  G. M. Dunn, A. B. Walker, A. J. Vickers, and V. R. Wicks, J. Appl. Phys. **79**(9), 7329-7338 (1996).

[44]  D. W. Snoke, W. W. Rühle, Y.-C. Lu, E. Bauser, Phys. Rev. Lett. **68**(7), 990–993 (1992).

[45]  S. M. Sze, K. K. Ng, Semiconductor Devices (John Wiley & Sons, Hoboken, New Jersey, 2007),  p. 789.

[46]  K. Kash, J. Shah, D. Block, A.C. Gossar, and W. Wiegmann, Physica **134**B, 189-198, (1985).

[47]  A. Gregory, S. Usher, F.A. Majumder, and R.T. Phillips, Solid State Commun. **87**(7), 605-608 (1993).

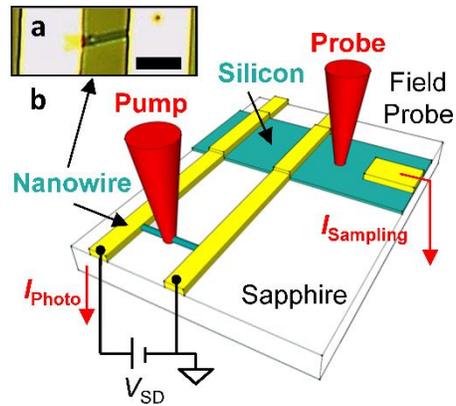

**Figure 1** Device geometry and optoelectronic pump-probe circuit. a) Optical microscopy image of an InAs-nanowire contacted by gold strip-lines. Scale bar: 4 µm. b) Schematic on-chip detection geometry. The pump laser pulse is focused on the InAs-nanowire contacted by the gold strip-lines. The probe pulse triggers the time resolved photocurrent signal $I_{Sampling}$ which is measured at the field probe. Gold electrodes are depicted in yellow, the nanowire and the ion implanted silicon in turquois.

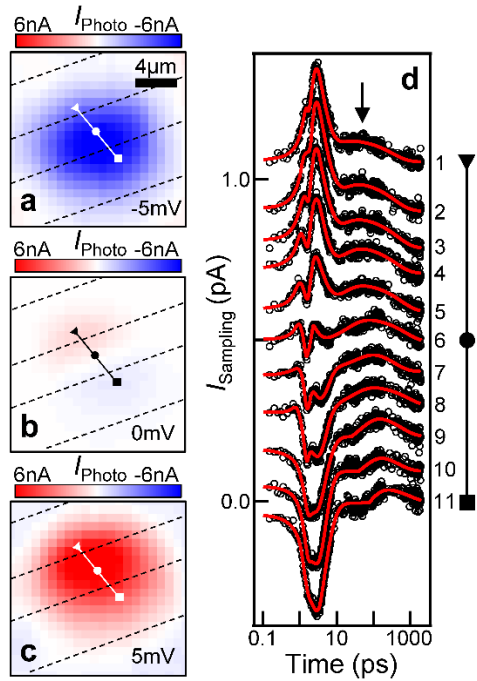

**Figure 2** Time integrated photocurrent $I_{Photo}$ of an InAs-nanowire measured for a) $V_{SD}$ = -5mV, b) $V_{SD}$ = 0V, and c) $V_{SD}$ = +5mV. Dashed lines indicate the position of the gold strip lines. Full line indicates position of the nanowire. The triangle and the square indicate the nanowire-gold contact position, and the circle highlights the center position in-between the strip lines. d) Time resolved photocurrent $I_{Sampling}$ for $V_{SD}$ = +5mV and different pump laser positions along the nanowire starting at the position 'triangle' (data on the top) to the position 'square' (data on the bottom). Lines are fits to the data. Arrow indicates the position of a peak due to the hole drift ($\lambda$ = 780 nm, $P_{Pump}$ = 6,4 kW/cm$^2$, $P_{Probe}$ = 3,2 MW/cm$^2$, $T_{Bath}$ = 77 K, sample #1).

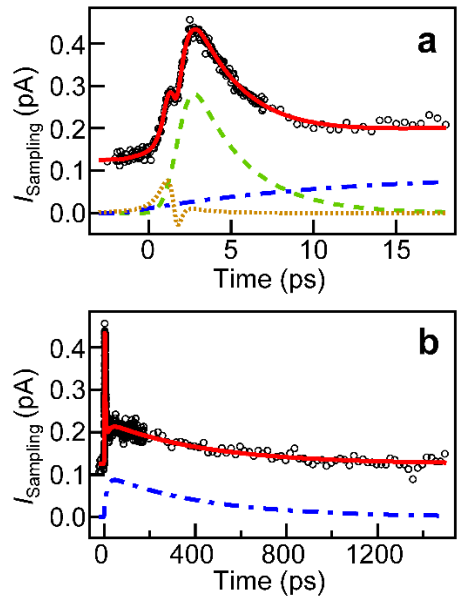

**Figure 3** Time resolved photocurrent $I_{\text{Sampling}}$ and individual fit components for position 3 in
Figure **2** on a time scale up to a) 20 ps and b) 1500 ps. See text for details.

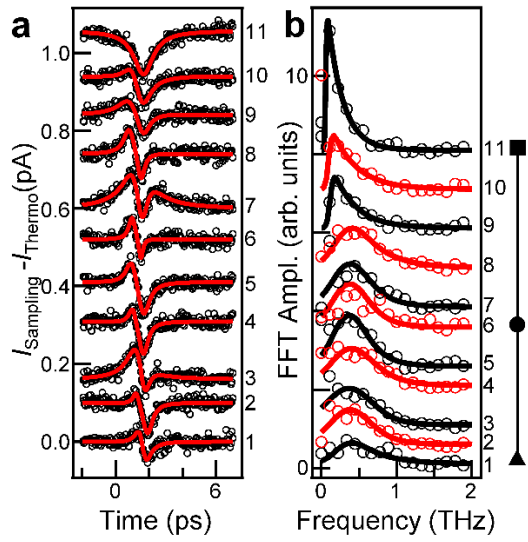

**Figure 4** THz component of $I_{\text{Sampling}}$ in
Figure **2**d plotted a) in the time domain and b) the frequency domain. Lines in a) are fits to
the THz component. Lines in b) are fits to the Fourier spectra.

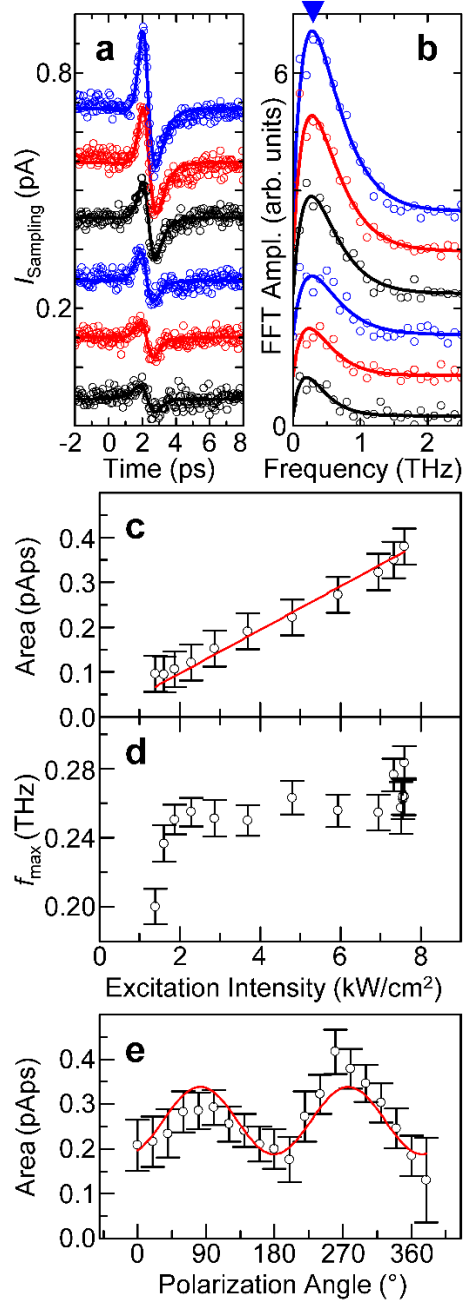

**Figure 5** Intensity-dependence of the THz oscillation. a) Time-resolved photocurrent $I_{Sampling}$ and fits for excitation intensities 1.4 kW/cm², 1.9 kW/cm², 3.7 kW/cm², 4.8 kW/cm², 6.9 kW/cm² and 7.6 kW/cm² (bottom to top). b) Corresponding Fourier spectra of the data with fit function. c) Area of the THz oscillation and d) oscillation frequency plotted over the excitation intensity of the pump laser pulse. e) Area of the THz oscillation as a function of the linear laser polarization with a sinusoidal fit ($V_{SD}$ = 0 V, $\lambda$ = 780 nm, $P_{Probe}$ = 3,2 MW/cm², $T_{Bath}$ = 77 K, sample #2).

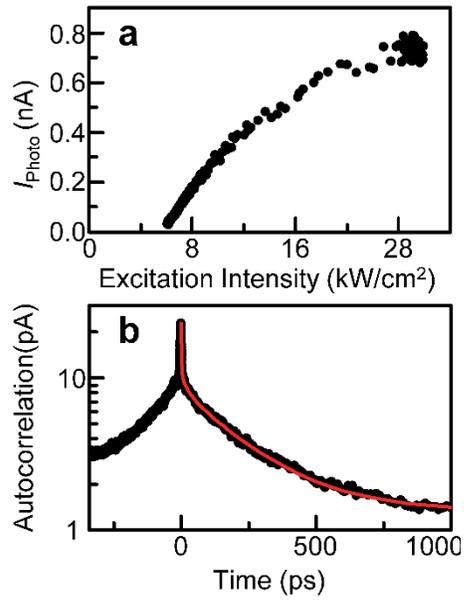

**Figure 6** $t_{\text{hole}}$ versus the distance of the pump laser pulse to the negative contact (top strip-line in Figure **2**c. Red line is a linear fit to the data. See text for details.

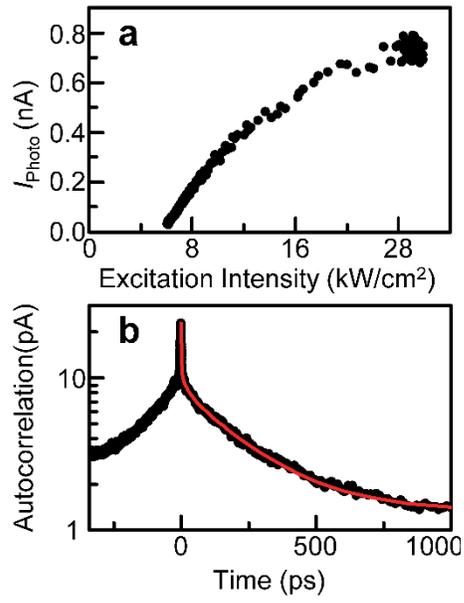

**Figure 7** a) Intensity-dependence of the time-integrated photocurrent $I_{Photo}$ at the position indicated with a square in

Figure **2**b. b) Photocurrent autocorrelation signal at this position. Line fits the data on the positive time scale with a tri-exponential decay ($\lambda = 780$ nm, $V_{SD} = 0$ V, $T_{Bath} = 77$ K, sample #1).